\documentclass[adp,fleqn]{w-art}
\usepackage{times,cite}
\usepackage{w-thm}
\usepackage[]{graphicx}
\chardef\bslash=`\\ 
\newcommand{\ntt}{\normalfont\ttfamily}
\newcommand{\cn}[1]{{\protect\ntt\bslash#1}}

\begin{document}
\DOIsuffix{theDOIsuffix}
\Volume{XX}
\Issue{X}
\Copyrightissue{XX}
\Month{MM}
\Year{2009}
\pagespan{1}{}
\Receiveddate{DD Month 2009}
\Accepteddate{DD Month 2009}
\keywords{Two-dimensional conductivity, spin accumulation, Si-MOSFET.}
\subjclass[pacs]{72.25.Mk,72.25.Dc,73.40.-c} 



\title[Conductance asymmetry \dots]{Conductance asymmetry of a slot gate 
Si-MOSFET in a strong parallel magnetic field}


\author[I. Shlimak]{I. Shlimak\footnote{Corresponding
     author \quad E-mail: {\sf shlimai@mail.biu.ac.il}, Phone: +972\,3\,531\,8176,
     Fax: +972\,3\,531\,7749}\inst{1}}
\address[\inst{1}]{Jack and Pearl Resnick Institute of Advanced Technology, 
	Department of Physics, Bar-Ilan University, Ramat-Gan 52900,  Israel}
\author[D. I. Golosov]{D. I. Golosov 
\inst{1}}

\author[A. Butenko]{A. Butenko 
	\inst{1}}
\author[K.-J. Friedland]{K.-J. Friedland 
	\inst{2}}
\address[\inst{2}]{Paul-Drude Institut f\" ur Festk\"orperelektronik, 
	Hausvogteiplatz 5-7, 10117, Berlin, Germany}
\author[S. V. Kravchenko]{S. V. Kravchenko 
	\inst{3}}
\address[\inst{3}]{Physics Department, Northeastern University, 
	Boston, Massachusetts 02115, U.S.A.}

\begin{abstract}
	We report measurements on a Si-MOSFET sample with a slot 
	in the upper gate, allowing for different electron densities $n_{1,2}$ 
	across the slot. The dynamic longitudinal resistance was measured 
        by the standard 
	lock-in technique, while maintaining
	a large DC current through the source-drain channel.  
	We find that the conductance of the sample in a strong parallel 
   magnetic field is asymmetric with respect to the DC current direction. 
	This asymmetry increases with  magnetic field. 
	The results are interpreted in terms of electron spin accumulation or 
        depletion near the slot.
\end{abstract}
\maketitle




\renewcommand{\leftmark}
{I. Shlimak et al.: Conductance asymmetry \dots}

\section{Introduction}
\label{sect1}

The objective of this work was to probe the influence of strong parallel 
magnetic field 
on electron transport across an interface between  regions with different 
electron densities, $n_1$ and $n_2$, in a single Si-MOSFET sample. The
sample has a narrow slot in the upper gate, which 
allows one to apply different voltages to separate gates. Previously, the
longitudinal conductivity  of a slot-gate Si-MOSFET sample was
measured in a
perpendicular magnetic field in the quantum Hall effect (QHE) regime
\cite{Shlimak}. For equal gate voltages, the presence of the slot did
not cause any measurable decrease in conductance, implying that
the slot does not act as a potential barrier for
electrons\cite{Shlimak}.

The effect of a parallel magnetic field on the 
conductance of a two-dimensional electron gas (2DEG) in spatially
uniform Si-MOSFET samples had been investigated 
earlier\cite{Simonian, Tsui, Shashkin} in the context of
metal-insulator transition studies. The conductance asymmetry with
respect to the direction of the electric current (always 
parallel or antiparallel to
the magnetic field), reported here, is a novel effect associated with
the non-uniform properties of our slot-gate sample. Phenomenological
interpretation of our results (involving current-induced spin
accumulation or depletion near the slot) suggests that this asymmetry 
is directly related to the physical mechanism underlying
the parallel-field magnetoresistance of a Si-MOSFET 2DEG.     

\section{Experiment}
\label{sect2}

The sample used in our experiments was investigated
earlier (see Ref. \cite{Shlimak}). 
A narrow slot ($100$~nm) had been made in the upper metallic gate, allowing
one to apply different 
gate voltages to different parts of the gate and thereby to
independently control the electron density 
in the two areas of the sample. Measuring  the transverse Hall
resistivity, 
$\rho_{xy}$, and longitudinal resistivity, $\rho_{xx}$, 
as functions of the gate voltage $U_{\rm G}$ in a perpendicular magnetic
field yields the dependence of the
electron density $n$ on $U_{\rm G}$: $n = 1.43\cdot 10^{15}\, (U_{\rm G}
- 0.64{\rm ~V}){\rm ~m}^{-2}$;  at $n = 1.62\cdot 10^{16}~{\rm
  m}^{-2}$, electron 
mobility equals 
$\mu = 1.46~{\rm m}^2/{\rm (V\cdot s)}$.

For the next series of measurements, the sample was mounted along the 
magnet axis, 
so that the current flow would be parallel to the magnetic field. 
The misalignment between the two
was
estimated with the help of Hall effect measurements. Whereas  
the Hall voltage must vanish in an ideal 
parallel geometry, the small value 
registered  corresponds to a minute misalignment of 
$\sim 0.1^{\rm o}$.

Our experimental scheme enables one to pass a DC current, $I_{\rm DC}$, 
of up to $1~\mu{\rm A}$ through the
source-drain 
channel, while measuring the dynamic resistance at  $12.7$~Hz 
via a standard lock-in technique with an AC current of 
$\sim 50$~nA. The sample temperature was maintained at $300$~mK. 

We fix the gate voltages at $U_{\rm G}(1) = 7{\rm ~V}$ (
corresponding
to $n_1 = 0.9\cdot 10^{16}{\rm ~m}^{-2}$ in area 1 of the sample) and 
$U_{\rm G}(2) = 18{\rm ~V}$ ($n_2 = 2.5\cdot 10^{16}{\rm ~m}^{-2}$ in area 2).
Fig. \ref{fig:data} (a) displays the conductance $\sigma$ of our sample
measured as a function of 
$I_{\rm DC}$ in the absence of a magnetic field, at 
$B = 7$~T, and  at $14$~T. 
One can see the following features: 

\begin{figure}[t]
\includegraphics[width=.9\textwidth]{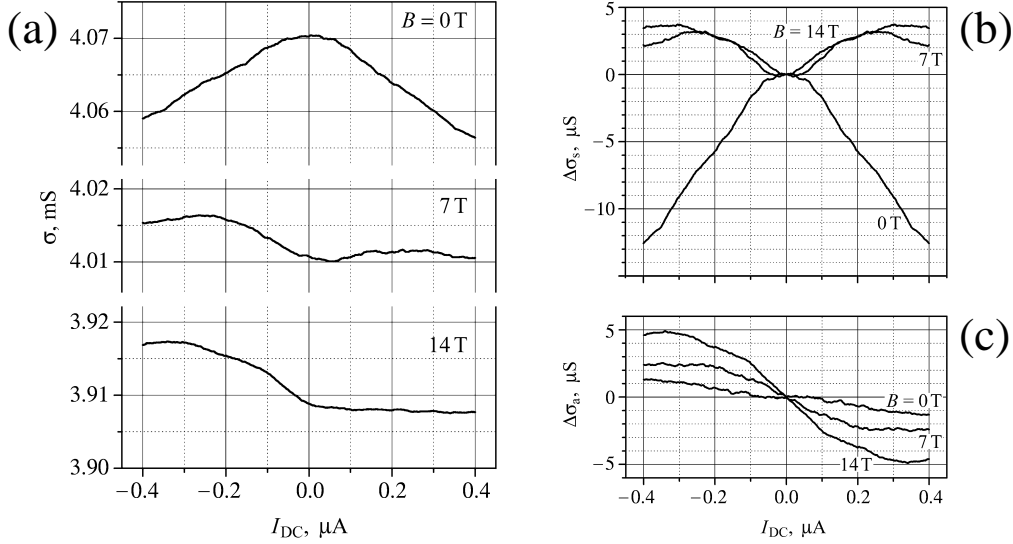}%
\caption{(a) Sample conductance as a function of DC current at $B = 0$, $7$, 
and  
$\pm14$~T . $U_{\rm G}(1) = 7$~V, $U_{\rm G}(2) = 18$~V. (b) and (c) show 
symmetric and asymmetric parts of $\sigma(I_{\rm DC})-\sigma(0)$.}
\label{fig:data}
\end{figure} 

1)	At zero $I_{\rm DC}$,  positive magnetoresistance (PMR) or negative 
magnetoconductance (NMC) is observed: the conductance decreases with 
increasing magnetic field.  The magnitude of 
NMC is [$\sigma (B=0) - \sigma (B)]/\sigma (B=0) = 1.5\,\%$ 
for $B = 7$~T, and  3.9\,\% for $B = 14$~T.

2)	At $B = 0$, the conductivity decreases slightly with the DC current, and 
$\sigma (I_{\rm DC})$ is almost symmetric with respect to the sign of  $I_{\rm DC}$.

3)	At $B = 7$ and 14~T, the dependence $\sigma (I_{\rm DC})$ is clearly asymmetric. 
This asymmetry does not depend on the direction of the magnetic field:
the shape of the curves 
is identical for $B = 14$~T and $ -14$~T. This excludes the Hall
voltage (which may arise due to the slight misalignment of the sample) 
as a possible origin of the asymmetry. 

\section{Discussion}
\label{sect3}

We will now discuss the observed behaviour of conductance in more detail.

\noindent 1. For Si-based two-dimensional systems, the NMC effect in a parallel
magnetic field had been reported 
earlier\cite{Simonian,Tsui,Shashkin,Okamoto5,Vitkalov,Lai,Okamoto8}. 
The metallic-like conductivity of Si MOSFETs decreases with an increasing
in-plane magnetic field and stabilises once 
the electrons are fully polarised \cite{Okamoto5,Vitkalov}. 

\noindent 2.	Figs. \ref{fig:data} (b) and \ref{fig:data} (c) show the 
decomposition 
of $\sigma (I_{\rm DC})$ 
into  symmetric $\sigma_{\rm s}$ and antisymmetric $\sigma_{\rm a}$
parts according to
 $\Delta \sigma_{\rm s} = [\sigma(I_{\rm DC})  +  \sigma(-I_{\rm
  DC})]/2 - \sigma(I_{\rm DC} = 0)$, and 
$\Delta\sigma_{\rm a}=[\sigma(I_{\rm DC})  -  \sigma(-I_{\rm
  DC})]/2$.

At $B=0$, the most likely source of
$\Delta \sigma_{\rm s}$   is 
the Joule 
heating caused by $I_{\rm DC}$. In our case, both electron concentrations $n_1$ and $n_2$ 
correspond to  metallic behaviour, with increasing 
temperature at 
$B = 0$ leading to a conductance decrease, $d\sigma /dT < 0$, which explains 
the 
experimental data. In a strong magnetic field, however, the conductivity of 
Si-MOSFETs 
does not depend on temperature \cite{Shashkin}, and heating by a DC current does 
not affect the conductance. We indeed see that in a field, 
values of $\Delta \sigma_s$ become much smaller. This is accompanied by a
growth of the antisymmetric part, $\Delta \sigma_a$ (and hence of the
overall asymmetry of $\Delta \sigma (I_{\rm DC})$).  

A small asymmetry  observed at $B = 0$ (about $2.5\cdot 10^{-4}$ of the
net conductance at maximal current) can be explained by an additional 
voltage 
bias $V_{\rm DC}$ induced by $I_{\rm DC}$:  $V_{\rm DC} = I_{\rm
  DC}/\sigma $. 
For our sample geometry, $V_{\rm DC}$ at $I_{\rm DC} =
0.4$~$\mu {\rm A}$ reaches 
$1$~mV which is, indeed, about $10^{-4}$ of the $U_{\rm G} = 7$~V. 
In MOSFETs, $V_{\rm DC}$ with an appropriate sign is added to the gate voltage.
This leads to a small increase or decrease (depending on the sign of 
$I_{\rm DC}$) 
of the electron density and hence to a change in $\Delta\sigma$.
We emphasise that this mechanism cannot possibly account for the much
more pronounced asymmetric behaviour of $\sigma (I_{\rm DC})$ found in the
presence of a strong magnetic field.

\noindent 3. The observed enhancement of asymmetric behaviour of 
conductance in a 
parallel field [see Fig. \ref{fig:data} (c)] can be understood in terms
of electron spin accumulation/depletion near the interface. Consider, 
{\it e.g.}, the case of $I_{\rm DC}>0$, corresponding to the flow of 
(appropriately spin-polarised) electrons from area 1 to area 2, where the
relative spin polarisation is smaller. Below, we argue that such a current
causes a local increase of spin polarisation in area 2 near the slot,
resulting in a decrease of the overall conductance. 
 
Within a simple
Drude approach, applying in-plane magnetic
and electric fields to a uniform strictly two-dimensional system gives
rise to the magnetisation, electric current, and spin current densities:
\begin{equation}
M_0\equiv \frac{1}{2}(n_\uparrow - n_\downarrow) =
\frac{m_* \nu g\mu_B B}{4 \pi \hbar^2}\,,\,\,\,\,
j= \frac{e^2 E \tau}{m_*} (n_\uparrow + n_\downarrow)\,,\,\,\,
s=-\frac{eE \tau}{2m_*}(n_\uparrow - n_\downarrow)\,.
\end{equation}
Here, $g$ is the Land\'{e} factor, $\nu=2$ the number of valleys, 
$e=|e|$ and $m_*$ are electron charge
and effective mass, and $n_\uparrow$ 
($n_\downarrow$) density of spin-up (spin-down) electrons. The momentum
relaxation time $\tau$ (assumed to be the same for both spin directions) is
much shorter\cite{Wilamowski} than the longitudinal spin relaxation time $T_1$, and we
will be faced with a situation in which the magnetisation $M$ deviates from its
equilibrium value $M_0$. It is easy to see that the relationship
$s=-jM/en$, where $n=n_\uparrow + n_\downarrow$,
persists in such a non-equilibrium case. 

\begin{figure}[t]
\sidecaption
\includegraphics[width=.45\textwidth]{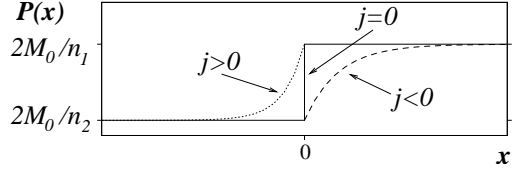}%
\caption{Schematic plot of the degree of spin polarisation, $P$, across
the interface. The solid, dotted and dashed lines correspond to $j=0$, $j>0$,
and $j<0$, respectively. The slot is located at  $x=0$.}
\label{fig:tails}
\end{figure}

The value $M_0$ does not depend on the electron density $n$. Therefore 
for $j=0$, 
at equilibrium, the quantity $P=2M/n$ (the degree of polarisation) suffers
a jump at the interface between the two parts of our sample [solid line in Fig.
\ref{fig:tails}, where we assume  $n_1<n_2$ with $n=n_1$ ($n=n_2$) for 
$x>0$ ($x<0$)].
For the measurement shown in Fig. \ref{fig:data}, the equilibrium value 
of $P$ at $B=14$~T can be estimated to be $P\approx 0.19$ for $n=n_1$ and 
$P \approx 0.07$ for $n=n_2$. 
For $j \neq 0$, the continuity of the electric and spin currents 
dictates that $P$ must also be continuous at $x=0$. Since the values of $n_1$ and
$n_2$ are fixed by the gate voltages, this in turn implies that $M$ must
deviate from $M_0$. Neglecting spin diffusion, we write for the steady
state:
\begin{equation}
-\frac{\partial s}{\partial x}\equiv \frac{j}{en} \frac{\partial  M}
{\partial x} = \frac{M(x)-M_0}{T_1} \,\,\,\,\Rightarrow
\,\,\,\,\,\frac{M(x)-M_0}{M_0}=\left\{ \begin{array}{ll}
( \frac{n_1}{n_2}-1 ) \theta(x) \exp(
\frac{en_1}{jT_1^{(1)}}x 
)\,,& j<0,
\\ 
& \\( \frac{n_2}{n_1}-1 ) \theta(-x) \exp(
\frac{en_2}{jT_1^{(2)}}x
)\,,
& j>0,
\end{array} \right.
\label{eq:M}
\end{equation}
with $\theta(x)=1$ for $x>0$ and $0$ otherwise; we also denote 
$T_1(n_{1,2})$ by $T^{(1,2)}_1$. Schematic profiles of $P(x)$, 
shown in Fig. \ref{fig:tails}, 
reflect accumulation (depletion)
of electron spin at $j>0$ ($j<0$). 

In a spatially uniform situation, two physical mechanisms are known to 
contribute to the dependence of the 
resistivity $\rho(n,B)$ on $B$: (i) the effect of spin-polarisation on the
screening properties\cite{Dolgopolov} and on electron correlations in an
interacting system\cite{Zala,Vitkalov03}, and (ii) orbital effects in the 
out-of-plane
direction\cite{Dassarma}. In the first case only, $\rho$ depends
on $B$ via the magnetisation $M_0(B)$; more precisely, $\rho$ depends not 
on $B$ but on $M$, 
regardless of whether the latter equals the equilibrium value, $M_0$. 
Denoting by $\alpha \leq 1$ the relative
contribution of this first mechanism, we may write
\begin{equation}
\rho(n,B,M)= \rho(n,B,M_0)+ \alpha \frac{\partial \rho(n,B)}{\partial B}
\left(\frac{d M_0}{dB} \right)^{-1}(M-M_0)\,,\,\,\,\,\rho(n,B,M_0) \equiv
\rho(n,B).
\label{eq:rho}
\end{equation}
Substituting Eqs. (\ref{eq:M}--\ref{eq:rho}) into the expression $V_{DC}=
j\int \rho[n,B,M(x)]dx$ for the bias voltage, after simple algebra we
find the differential conductance:
\begin{equation}
\sigma(I_{\rm DC},B)=\sigma(0,B)-\frac{2I_{\rm DC}B}{ed^2}\alpha [\sigma(0,B)]^2 
\left(\frac{1}{n_1}-\frac{1}{n_2}\right)\times 
\left\{ \begin{array}{ll} T_1^{(1)} \partial \rho(n_1,B)/\partial B\,,
  & I_{\rm DC}<0,
\\ 
& \\ T_1^{(2)} \partial \rho(n_2,B)/\partial B\,,
& I_{\rm DC}>0
\end{array} \right.
\nonumber
\end{equation}
where $d\sim 30$ $\mu {\rm m}$ is the width of our sample.
We see that $\sigma(I_{\rm DC},H)-\sigma(0,H)$ is proportional to $B$ and indeed 
changes sign at 
$I_{\rm DC}=0$, with different slopes for 
positive and negative $I_{\rm DC}$. Very rough estimates
suggest 
that the
value of the coefficient $\alpha$ may be about $\alpha  \stackrel{>}{\sim}
0.5$. 
Further
theoretical and experimental work is clearly needed to justify the many 
assumptions made in this preliminary treatment.

\begin{acknowledgement}
We thank A. Belostotsky and A. Bogush  for assistance. Discussions with
R. Berkovits, B. D. Laikhtman, and L. D. Shvartsman are gratefully 
acknowledged. This work was
supported by the BSF grant no 2006375 and by the Israeli Absorption Ministry. 
I. S. thanks the Erick and Sheila 
Samson Chair of Semiconductor Technology for financial support. 
\end{acknowledgement}

\end{document}